\documentclass[pre,twocolumn,aps,10pt]{revtex4-2}
\usepackage{amsmath}
\usepackage{amssymb}
\usepackage{graphicx}
\usepackage{braket}
\usepackage{stmaryrd}
\usepackage{hyperref}
\usepackage{color}
\usepackage{dsfont}
\usepackage{bm}

\hypersetup{
    colorlinks,
    citecolor=blue,
    linkcolor=blue,
    urlcolor=blue
}

\newcommand{\dblbrace}[1]{\llbracket #1\rrbracket}

\begin{document}

\title{Anisotropic Information Geometry of Entropy Production
}
\author{Tomohiro Nishiyama}
\email{htam0ybboh@gmail.com}
\affiliation{Independent Researcher, Tokyo 206-0003, Japan}

\author{Yoshihiko Hasegawa}
\email{hasegawa@biom.t.u-tokyo.ac.jp}
\affiliation{Department of Information and Communication Engineering, Graduate
School of Information Science and Technology, The University of Tokyo,
Tokyo 113-8656, Japan}

\date{\today}
\begin{abstract}

We reveal that the information geometry of entropy production exhibits an intrinsic anisotropy, providing a unified origin for the inequivalent geometric constraints obeyed by its different components. Using a hierarchical projection structure of quantum relative entropy, we derive an exact orthogonal decomposition of entropy production into three distinct geometric contributions associated with the deviation of the Gibbs projection temperature from the reference temperature, the deviation of the environment from its Gibbs projection state, and system--environment correlations. We demonstrate that these contributions obey fundamentally different geometric bounds: the first reduces to a classical Kullback--Leibler divergence and admits no universal upper bound in terms of the trace distance and the system and environment dimensions alone, whereas the latter two admit rigorous distance-based bounds with dimension-dependent factors. The framework also applies to classical Markovian dynamics described solely in terms of the system state, for which the system--environment correlation contribution is absent from the decomposition. Combining this geometric decomposition with dynamical speed limits, we obtain rigorous bounds on entropy production expressed solely in terms of physically accessible quantities, such as the mean energy and Hamiltonian variance.
\end{abstract}
\maketitle

\section{Introduction\label{sec:Introduction}}

Entropy production has been decomposed into various physically meaningful contributions in several contexts, including system--environment correlations and environmental nonequilibrium~\cite{Esposito:2010:EntProd}. While these decompositions have provided valuable physical insights, the geometric constraints governing each individual contribution and their quantitative relations with state-space distances have remained only partially understood. Previous studies have established bounds for specific contributions, such as the correlation contribution~\cite{Ptaszynski:2019:DominantEP}, highlighting their distinct physical roles. However, a unified geometric framework that comprehensively clarifies these diverse constraints remains lacking. In particular, it is still unexplored whether the distinct mathematical inequalities obeyed by these individual contributions can be explained by a deeper, overarching geometric principle underlying nonequilibrium thermodynamics.

In this work, we demonstrate that the information geometry of entropy production exhibits an intrinsic anisotropy. Here, anisotropy refers to the fact that different orthogonal components of entropy production are governed by inequivalent geometric constraints, despite originating from the same quantum relative entropy. Our approach is based on a hierarchical projection structure of quantum relative entropy onto nested statistical manifolds and builds upon the information-geometric Gibbs projection introduced in Ref.~\cite{reeb2014improved}. Successive applications of the Pythagorean theorem for quantum relative entropy yield an exact orthogonal decomposition of entropy production into three contributions associated with the deviation of the Gibbs projection temperature from the reference temperature, the deviation of the environment from its Gibbs projection state, and system--environment correlations. The Gibbs projection temperature is defined such that the corresponding Gibbs state has the same mean energy as the environment state.

We reveal that these three contributions exhibit fundamentally different relations with state-space geometry. The contribution associated with the deviation of the Gibbs projection temperature from the reference temperature reduces to a classical Kullback--Leibler divergence and is governed by a qualitatively distinct geometric structure: unlike the other two contributions, it cannot be universally upper-bounded by any function of the trace distance and the dimensions of the system and environment alone. In contrast, the contribution associated with the deviation of the environment from its Gibbs projection state and the system--environment correlation admit rigorous bounds in terms of their respective trace distances and dimension-dependent factors. These results show that entropy production is not geometrically isotropic: different directions obey different mathematical inequalities and, consequently, different information-theoretic constraints. This anisotropic structure also has direct dynamical consequences. Quantum speed limits provide a natural connection between state-space geometry and the rate of state evolution~\cite{deffner2020quantum}. While such bounds have been used to constrain entropy changes, our geometric decomposition reveals additional structure in entropy production by resolving its mean-energy contribution into a difference involving nonnegative quantities. This structure allows the distinct geometric constraints on the individual contributions to be translated into rigorous upper and lower bounds on entropy production.

As an illustrative case, consider classical Markovian dynamics, where the system state alone provides a closed description of the dynamics. In this setting, the system--environment correlation contribution vanishes. If the dynamics additionally satisfy detailed balance with respect to a single heat bath, the decomposition further reduces to a two-component structure governed by the total variation distance. This allows analogous bounds to be obtained using classical speed limits based on the dynamical activity~\cite{Hasegawa:2023:BulkBoundaryBoundNC}.

Moreover, the persistence of this anisotropic structure in systems coupled to multiple heat baths demonstrates that it originates from the underlying information geometry rather than from a particular thermal environment. Our framework, therefore, provides a unified geometric perspective connecting information geometry, thermodynamics, and dynamical speed limits.

\section{Methods}
\subsection{Pythagorean identity}

Throughout this paper, we set $\hbar=k_B=1$.
For density operators $\rho$ and $\sigma$, the quantum relative entropy is defined as 
\begin{align}
    D(\rho\Vert\sigma):=\mathrm{tr}[\rho(\ln \rho-\ln\sigma)].
    \label{eq:def_quantum_RE}
\end{align}
We define the von Neumann entropy as
\begin{align}
    S(\rho):=-\mathrm{tr}[\rho \ln \rho].
\end{align}
In this paper, we say that $\rho$, $\sigma$, and $\omega$ are orthogonal if they satisfy
\begin{align}
    \mathrm{tr}[(\rho-\sigma)(\ln\sigma-\ln\omega)]=0.
    \label{eq:orthogonal_cond}
\end{align}
This condition expresses the dual affine (Bregman) orthogonality associated with the information-geometric projection of $\rho$ onto $\omega$ in the space of density operators~\cite{amari2000methods, csiszar1975divergence},
and is equivalent to the optimality condition that yields the Pythagorean theorem for the quantum relative entropy:
\begin{align}
    D(\rho\Vert\omega)=D(\rho\Vert\sigma)+D(\sigma\Vert\omega).
    \label{eq:Pythagorean_ID}
\end{align}
The details of the derivations are shown in Appendix~\ref{sec:pythagorean}.
Let $H$ be a Hamiltonian with at least two distinct eigenvalues with arbitrary degeneracies.
For an inverse temperature $\beta$, let us define the Gibbs state as 
\begin{align}
    \gamma(\beta):=\frac{e^{-\beta H}}{\mathcal{Z}(\beta)},
    \label{eq:def_Gibbs_state}
\end{align}
where $\mathcal{Z}(\beta):=\mathrm{tr}[e^{-\beta H}]$ denotes the partition function.
We define the \textit{Gibbs projection inverse temperature} $\beta^{\ast}$ via the energy-matching condition:
\begin{align}
    \mathrm{tr}[H\gamma(\beta^{\ast})]=\mathrm{tr}[H\rho].
    \label{eq:def_eff_beta}
\end{align}
For $\beta^{\ast} \in (-\infty, \infty)$, this value is uniquely determined (see Appendix~\ref{sec:uniqueness}).
Equation~\eqref{eq:orthogonal_cond} implies the orthogonality of $\rho$, $\gamma(\beta^{\ast})$, and $\gamma(\beta)$, from which the Pythagorean identity follows~\cite{reeb2014improved}:
\begin{align}
    D(\rho \Vert \gamma(\beta))=D(\rho \Vert \gamma(\beta^{\ast}))+D(\gamma(\beta^{\ast})\Vert\gamma(\beta)).
    \label{eq:Pythagorean_eff}
\end{align}
From the result in Ref.~\cite{reeb2014improved}, it follows that 
\begin{align}
    &D(\rho \Vert \gamma(\beta^{\ast}))=-S(\rho)+\beta^{\ast} \mathrm{tr}[H\gamma(\beta^{\ast})]+\ln \mathcal{Z}(\beta^{\ast})\nonumber\\
    &= S(\gamma(\beta^{\ast}))-S(\rho),
    \label{eq:dif_entropy}
\end{align}
where we use Eq.~\eqref{eq:def_eff_beta} in the first equality and $D(\gamma(\beta^{\ast})\Vert \gamma(\beta^{\ast}))=0$ in the second equality.
Let $\|\rho-\sigma\|_1:=\mathrm{tr}[\sqrt{(\rho-\sigma)^\dagger (\rho-\sigma)}]$ denote the trace norm, and let $\mathcal{T}(\rho,\sigma):=\|\rho-\sigma\|_1/2$ be the trace distance. For the difference in von Neumann entropy, the Fannes–Audenaert inequality~\cite{fannes1973continuity,audenaert2007sharp} provides an upper bound in terms of the trace distance:
\begin{align}
    |S(\rho)-S(\sigma)|\le \mathcal{T}(\rho,\sigma)\ln (d-1)+h_{b}(\mathcal{T}(\rho,\sigma)),
    \label{eq:FA_bound}
\end{align}
Here, $d$ denotes the dimension of the Hilbert space supporting $\rho$ and $\sigma$, and $h_{b}(p):=-p\ln p-(1-p)\ln(1-p)$ is the binary entropy for $0\le p\le 1$. The upper bound is attained if $\rho$ and $\sigma$ can be simultaneously diagonalized in a common basis such that their eigenvalues take the form:
\begin{align}
    \rho&=\mathrm{diag}\left(1-\mathcal{T}(\rho,\sigma), \frac{\mathcal{T}(\rho,\sigma)}{d-1}, \ldots, \frac{\mathcal{T}(\rho,\sigma)}{d-1}\right), \nonumber\\
    \sigma &=\mathrm{diag}(1,0,\ldots, 0).
\end{align}

These results reduce to their classical counterparts when $\rho$, $\sigma$, and $\omega$ are simultaneously diagonalizable in the energy eigenbasis $\{\ket{\epsilon_i}\}$ of the Hamiltonian, such that $\rho=\sum_i p_i \ket{\epsilon_i}\bra{\epsilon_i}$. In this case, the quantum relative entropy and the von Neumann entropy reduce to the Kullback-Leibler (KL) divergence and the Shannon entropy, respectively:
\begin{align}
D_{\mathrm{KL}}(P\Vert Q):=\sum_i p_i \ln \frac{p_i}{q_i}, \label{eq:def_KL}
\end{align}
\begin{align}
H_{\mathrm{SH}}(P):=-\sum_i p_i \ln p_i.
\end{align}
Furthermore, the trace distance simplifies to the total variation distance $d_{\mathrm{TV}}(P,Q):=\frac{1}{2}\sum_i |p_i-q_i|$. Consequently, the bound in Eq.~\eqref{eq:FA_bound} reduces to its classical analog:
\begin{align}
|H_{\mathrm{SH}}(P)-H_{\mathrm{SH}}(Q)|\le d_{\mathrm{TV}}(P,Q)\ln (d-1)+h_{b}(d_{\mathrm{TV}}(P,Q)), \label{eq:FA_bound_classic}
\end{align}
for which the condition for equality remains analogous to the quantum case.
Under the same representation, the quantum Gibbs state naturally reduces to its classical counterpart, the Gibbs distribution $G(\beta) := \{\mathfrak{g}_i(\beta)\}$, defined as
\begin{align}
    \mathfrak{g}_i(\beta) := \frac{e^{-\beta \epsilon_i}}{\mathcal{Z}(\beta)},
    \label{eq:def_Gibbs_dist}
\end{align}
where $\mathcal{Z}(\beta) := \sum_i e^{-\beta \epsilon_i}$ is the classical partition function.
Accordingly, the definition of $\beta^{\ast}$ simplifies to 
\begin{align}
\sum_i \epsilon_i p_i = \sum_i \epsilon_i \mathfrak{g}_i(\beta^{\ast}).
\label{eq:def_eff_beta_classic}
\end{align}

\begin{figure}
\centering
\includegraphics[width=1\linewidth]{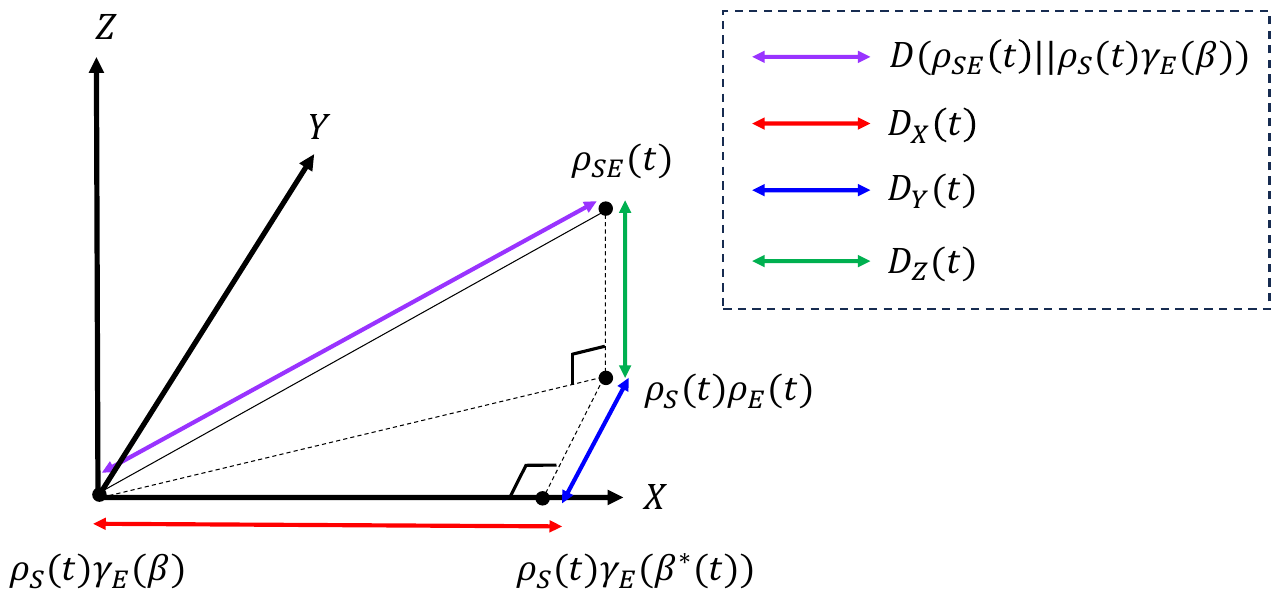}
\caption{Geometric decomposition of $D(\rho_{SE}(t)\Vert\rho_{S}(t)\gamma_E(\beta))$ in an open quantum system. The entropy production is expressed as the difference in $D(\rho_{SE}(t)\Vert\rho_{S}(t)\gamma_E(\beta))$ between distinct time. The definitions of $D_{X}(t)$, $D_{Y}(t)$ and $D_{Z}(t)$, which correspond to metrics of each axis, are provided in Eqs.~\eqref{eq:def_X_dist}--\eqref{eq:def_Z_dist}. 
}
\label{fig:geometry_EP_Q}
\end{figure}

\begin{figure}
\centering
\includegraphics[width=1\linewidth]{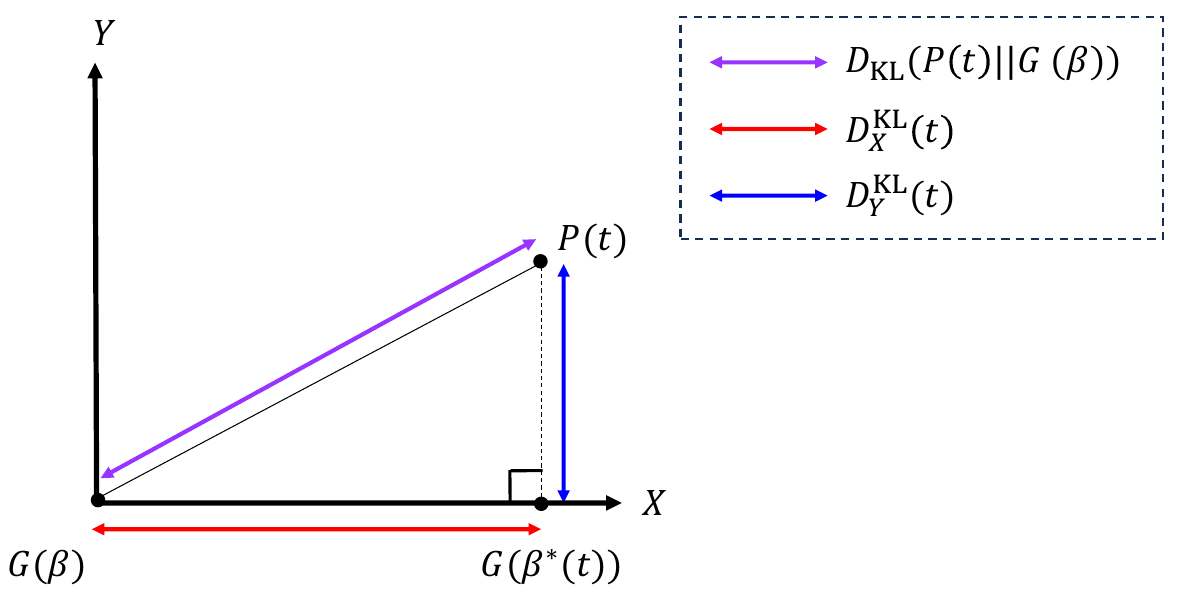}
\caption{Geometric decomposition of $D_{\mathrm{KL}}(P(t)\Vert G_(\beta))$ in a classical Markov system. The entropy production is represented by the difference in $D_{\mathrm{KL}}(P(t)\Vert G_(\beta))$ between different times. In the classical Markov regime, the $Z$ axis, representing system-environment correlations, is absent. The definitions of $D^{\mathrm{KL}}_{X}(t)$ and $D^{\mathrm{KL}}_{Y}(t)$, which correspond to metrics of each axis, are provided in  Eqs.~\eqref{eq:def_X_dist_C} and~\eqref{eq:def_Y_dist_C}. 
}
\label{fig:geometry_EP_C}
\end{figure}

\section{Results}
\subsection{Open quantum system}

Consider an open quantum system composed of a system $S$ and an environment $E$ with dimensions $d_S < \infty$ and $d_E < \infty$, respectively. 
Let $H_S(t)$ and $H_E$ denote the Hamiltonian operators of the system and the environment. The interaction between the two is governed by the time-dependent interaction Hamiltonian $H_{SE}(t)$. We assume that $H_E$ is time-independent and possesses at least two distinct eigenvalues. The total Hamiltonian is given by $H(t) = H_S(t) + H_E + H_{SE}(t)$.
Let $\rho_{SE}(t)$ represent the joint density operator of the system and environment. The time evolution of $\rho_{SE}(t)$ is governed by the total unitary evolution operator $U_t = \mathbb{T} e^{-i \int_0^t H(s) ds}$ such that
\begin{align}
    \rho_{SE}(t)=U_{t}\rho_{SE}(0) U_{t}^\dagger.
    \label{eq:evolution_density}
\end{align}
Here $\mathbb{T}$ denotes the time-ordering operator.
Crucially, we do not assume any specific form for the initial state $\rho_{SE}(0)$, such as a product state or a Gibbs state.  Let $\mathrm{tr}_{X}$ be a partial trace with respect to $X:=\{S,E\}$, and let $\rho_{S}(t):=\mathrm{tr}_{E}[\rho_{SE}(t)]$ and $\rho_E(t):=\mathrm{tr}_{S}[\rho_{SE}(t)]$ be the density operators of the system and the environment, respectively. 

Let $\gamma_{E}(\beta)$ be the Gibbs state for $H_{E}$ defined in Eq~\eqref{eq:def_Gibbs_state}.
Throughout this section, $\beta$ denotes an arbitrary but fixed inverse-temperature parameter defining the reference Gibbs state $\gamma_{E}(\beta)$.
Following the construction in Eq.~\eqref{eq:def_eff_beta}, the Gibbs projection temperature for the open quantum dynamics is defined as
\begin{align}
    \mathrm{tr}_{E}[H_E\gamma_E(\beta^{\ast}(t))]=\mathrm{tr}_{E}[H_{E}\rho_{E}(t)].
\label{eq:def_eff_beta_quantum}
\end{align}
The entropy production from time $t=0$ to $\tau$ is expressed as~\cite{Esposito:2010:EntProd, riechers2021initial, mondal2023modified, ptaszynski2023ensemble}.
\begin{align}
    &\Sigma(\tau)\nonumber\\
    &=D(\rho_{SE}(\tau)\Vert\rho_{S}(\tau)\otimes\gamma_E(\beta))-D(\rho_{SE}(0)\Vert\rho_{S}(0)\otimes\gamma_E(\beta)).
    \label{eq:EP_relative_entropy}
\end{align}
In Eq.\eqref{eq:EP_relative_entropy}, we incorporate an additional term to account for initial correlations, which was omitted in the original formulation in Ref.~\cite{Esposito:2010:EntProd}. This revised expression is consistent with Eq.~(4) in Ref.\cite{mondal2023modified}.
In the following, we omit the tensor product symbol $\otimes$ for simplicity. 

Let us focus on the term $D(\rho_{SE}(t)\Vert\rho_{S}(t)\gamma_E(\beta))$ on the right-hand side of Eq.~\eqref{eq:EP_relative_entropy}. 
As shown in Ref.~\cite{Esposito:2010:EntProd}, the entropy production can be decomposed into two distinct non-negative contributions:
\begin{align}
    D(\rho_{SE}(t)\Vert\rho_{S}(t)\gamma_E(\beta)) = I_{S:E}(t) + D(\rho_E(t) \Vert \gamma_E(\beta)), \label{eq:EP_decomp_old}
\end{align}
where $I_{S:E}(t)$ denotes the quantum mutual information between the system and the environment at time $t$:
\begin{align}
I_{S:E}(t) &:= S(\rho_S(t)) + S(\rho_E(t)) - S(\rho_{SE}(t))\nonumber\\
 &= D(\rho_{SE}(t)\Vert\rho_{S}(t)\rho_E(t)) \geq 0. \label{eq:def_mutual_info}
\end{align}
By applying Eq.~\eqref{eq:Pythagorean_eff}, the decomposition in Eq.~\eqref{eq:EP_decomp_old} can be further refined as follows:
\begin{align}
    &D(\rho_{SE}(t)\Vert\rho_{S}(t)\gamma_E(\beta))=D(\gamma_E(\beta^{\ast}(t))\Vert\gamma_E(\beta))
    \nonumber\\
    &+D(\rho_E(t)\Vert\gamma_E(\beta^{\ast}(t)))+D(\rho_{SE}(t)\Vert\rho_{S}(t)\rho_E(t)).
    \label{eq:EP_decomposition}
\end{align}
The geometric interpretation of this decomposition is illustrated in Fig.~\ref{fig:geometry_EP_Q}, where the meaning of each axis is described below:
\begin{itemize}
    \item \textbf{$X$-axis:}  The deviation of the Gibbs projection temperature from the reference temperature.
    \begin{align}
        D_{X}(t)&:=D(\gamma_E(\beta^{\ast}(t))\Vert\gamma_E(\beta)). \label{eq:def_X_dist}
    \end{align}
    \item \textbf{$Y$-axis:}  The deviation of the environment from its Gibbs projection state. 
    \begin{align}
        D_{Y}(t)&:=D(\rho_E(t)\Vert\gamma_E(\beta^{\ast}(t))).
        \label{eq:def_Y_dist}    \end{align}
    \item \textbf{$Z$-axis:} The deviation from the product state, quantifying system-environment correlations.
     \begin{align}
        D_{Z}(t)&:=D(\rho_{SE}(t)\Vert \rho_{S}(t)\rho_{E}(t)).
        \label{eq:def_Z_dist}
    \end{align}
\end{itemize}

Here, we use $D(\rho_{S}\sigma_{E}\Vert \rho_{S}\sigma^\prime_{E})=D(\sigma_{E}\Vert \sigma^\prime_{E})$ in Fig.~\ref{fig:geometry_EP_Q}.
Following the construction of Eqs.~\eqref{eq:def_X_dist}--\eqref{eq:def_Z_dist}, we analogously define trace distances $\mathcal{T}_{X}(t)$, $\mathcal{T}_{Y}(t)$, and $\mathcal{T}_Z(t)$. 
These three axes correspond to orthogonal components of entropy production.
The resulting decomposition reveals an intrinsic anisotropy in the information geometry underlying entropy production. Here, anisotropy refers to the fact that different directions in the decomposed entropy-production space are subject to inequivalent geometric constraints, even though they originate from the same quantum relative entropy. In other words, the geometry governing entropy production is direction-dependent, and different axes may exhibit fundamentally different relations with the underlying state-space structure. The following properties characterize this anisotropic behavior:
\begin{description}
    \item[Property 1 ($X$-axis)] 
    The quantity $D_{X}(t) = D_{\mathrm{KL}}(G_E(\beta^{\ast}(t)) \Vert G_E(\beta))$ represents a classical contribution that depends solely on the mean energy and temperature, rather than explicitly on the density operator. Crucially, $D_{X}(t)$ cannot be upper-bounded merely by the trace distance $\mathcal{T}_{X}(t)$ and the system dimensions; that is, there exists no function $f$ such that $D_{X}(t) \le f(\mathcal{T}_{X}(t); d_S, d_E)$. Consequently, $D_{X}(t)$ can take arbitrarily large values even in the regime where $\mathcal{T}_{X}(t) \ll 1$. Conversely, $D_{X}(t)$ can be lower-bounded by the trace distance as follows:
    \begin{align}
        2\mathcal{T}_{X}(t)^2\le D_{X}(t).
        \label{eq:l_bound_X}
    \end{align}
    
    \item[Property 2 ($Y$-axis)] Unlike $D_{X}(t)$, the quantity $D_{Y}(t)$ depends explicitly on the density operator of the environment. Consequently, it admits rigorous upper and lower bounds determined by the trace distance and the dimension of the environment $d_E$, as given by
     \begin{align}
        2\mathcal{T}_{Y}(t)^2 &\le D_{Y}(t) \le \mathcal{T}_{Y}(t)\ln(d_E-1)+h_{b}\left(\mathcal{T}_{Y}(t)\right).
        \label{eq:lu_bound_Y}
    \end{align}

    \item[Property 3 ($Z$-axis)] The quantity $D_{Z}(t)$ also explicitly depends on the density operator and admits rigorous upper and lower bounds determined by the trace distance and the smaller dimension of the system and environment, $\min(d_S,d_E)$, as given by
    \begin{align}
        2\mathcal{T}_{Z}(t)^2 &\le D_{Z}(t) \le 2\mathcal{T}_{Z}(t)\min(\ln d_{S}, \ln d_{E})+g\left(\mathcal{T}_{Z}(t)\right),
        \label{eq:lu_bound_Z}
    \end{align}
    where 
    \begin{align}
        g(x)&:=(1+x)h_{b}\left(\frac{x}{1+x}\right)\nonumber\\
        &=(x+1)\ln(x+1)- x\ln x.
    \end{align}

\end{description}

Note that the trace distances $\mathcal{T}_{Y}(t)$ and $\mathcal{T}_{Z}(t)$ lie in the interval $[0,1]$.
Furthermore, the corresponding entropy contributions satisfy $D_{Y}(t)\le \ln d_E$ and $D_{Z}(t)\le 2\min(\ln d_S, \ln d_E)$, where the latter bound follows from the Araki--Lieb inequality~\cite{araki1970entropy} (see Appendix~\ref{sec:proof_properties} for details).
Properties 1, 2, and 3 constitute the main results of this paper. 

While the lower bounds in Eqs.~\eqref{eq:l_bound_X}--\eqref{eq:lu_bound_Z} follow from the quantum Pinsker inequality~\cite{OhyaPetz:2004:QuantumEntropy}, the remaining proofs of Properties 1--3 are given in Appendix~\ref{sec:proof_properties}.
Since the range of $D_Y(t)$ is bounded by the environment dimension $d_E$, whereas that of $D_Z(t)$ is limited by $\min(d_S, d_E)$, the set of admissible values for $D_Z(t)$ is more strongly constrained by the dimensional structure of the system and the environment. 

The decomposition shown in Eq.~\eqref{eq:EP_decomposition} can be viewed as a decomposition with respect to a resource monotone. 
Let $\mathcal{F}$ be the set of free states, which are states that have no resource. 
Free operations constitute a set of operations that map a free state to a free state.
Let $\mathcal{E}$ be a free operation, which is a completely positive trace-preserving (CPTP) map. 
Then, free operations satisfy
\begin{align}
    \rho\in\mathcal{F}\Rightarrow\mathcal{E}(\rho)\in\mathcal{F}.
    \label{eq:free_state_def}
\end{align}
Given the free operations, we can define a resource monotone $R$ such that
\begin{align}
   R(\mathcal{E}(\rho))\le R(\rho).
   \label{eq:resource_monotone_def}
\end{align}
Moreover, the resource monotone is non-negative and is zero for free states:
\begin{align}
    \rho \in \mathcal{F} \Rightarrow R(\rho)=0.
    \label{eq:faithfulness}
\end{align}
The relative entropy is often employed as a resource monotone:
\begin{align}
    R_{\mathrm{rel}}(\rho)=\min _{\sigma \in \mathcal{F}} D(\rho \Vert \sigma).
    \label{eq:relative_entropy_monotone}
\end{align}
Using the monotonicity of the relative entropy, $R_\mathrm{rel}$ satisfies the condition of a resource monotone:
\begin{align}
R_{\mathrm{rel}}(\rho)&=\min_{\sigma\in\mathcal{F}}D(\rho\Vert\sigma)\nonumber\\&\ge\min_{\sigma\in\mathcal{F}}D(\mathcal{E}(\rho)\Vert\mathcal{E}(\sigma))\nonumber\\&=\min_{\sigma\in\mathcal{F}}D(\mathcal{E}(\rho)\Vert\sigma)=R_{\mathrm{rel}}(\mathcal{E}(\rho))
    \label{eq:relative_entropy_monotone2}
\end{align}
Let us define a set of free states $\mathcal{P}_{SE}$, which comprises the product states in $S$ and $E$:
\begin{align}
    \mathcal{P}_{SE}\equiv\{\rho_{S}\otimes\rho_{E}|\rho_{S}\in\mathcal{S}_{S},\rho_{E}\in\mathcal{S}_{E}\},
    \label{eq:P_SE_def}
\end{align}
where $\mathcal{S}_S$ and $\mathcal{S}_E$ are the sets of density operators in $S$ and $E$, respectively. 
Then, it can be shown that $D_Z(t)$ admits the following representation:
\begin{align}
    D_{Z}(t)&=\min_{\sigma_{SE}\in \mathcal{P}_{SE}}D(\rho_{SE}(t)\Vert\sigma_{SE}).
    \label{eq:DZ_variational}
\end{align}
Since Eq.~\eqref{eq:orthogonal_cond} is satisfied for $\rho=\rho_{SE}(t)$, $\sigma=\rho_S(t)\rho_E(t)$, and $\omega=\sigma_{SE}\in \mathcal{P}_{SE}$, we obtain
\begin{align}
    &D(\rho_{SE}(t)\Vert\sigma_{SE}) \nonumber\\
    &= D(\rho_{SE}(t)\Vert\rho_{S}(t)\rho_{E}(t)) + D(\rho_{S}(t)\rho_{E}(t) \Vert \sigma_{SE}), 
\end{align}
which gives Eq.~\eqref{eq:DZ_variational}.
Moreover, let $\mathcal{G}_E$ be the set of thermal states in $E$. 
Then, from Eq.~\eqref{eq:Pythagorean_eff}, $D_Y(t)$ in Eq.~\eqref{eq:def_Y_dist} can be expressed as
\begin{align}
    D_{Y}(t)&=\min_{\sigma_{E}\in \mathcal{G}_E}D(\rho_{E}(t)\Vert\sigma_{E}).
    \label{eq:DY_variational}
\end{align}
Equations~\eqref{eq:DZ_variational} and~\eqref{eq:DY_variational} show that 
$D_Z$ and $D_Y$ can be identified as resource monotones, where the corresponding free states are product states and thermal states, respectively.

By combining Eqs.~\eqref{eq:EP_relative_entropy} and~\eqref{eq:EP_decomposition} with the definitions in Eqs.~\eqref{eq:def_X_dist}--\eqref{eq:def_Z_dist}, it follows that
\begin{align}
    \Sigma(\tau)=\Delta D_{X}+\Delta D_{Y}+\Delta D_{Z}.
    \label{eq:EP_decomposition_diff}
\end{align}
Here, we denote a change by $\Delta F := F(\tau) - F(0)$ for an arbitrary time-dependent quantity $F(t)$.
The construction is presented for a single heat bath for clarity. The extension to multiple reservoirs is straightforward and is summarized in Appendix~\ref{sec:multi_bath}.
While the present discussion is based on the canonical ensemble, the arguments extend straightforwardly to the grand canonical ensemble. In this case, the reference state in the definition of entropy production [Eq.~\eqref{eq:EP_relative_entropy}] is replaced by the grand-canonical Gibbs state $\gamma_{E}(\beta, \mu) = \Xi_E^{-1}(\beta, \mu) e^{-\beta(H_{E}-\mu N_{E})}$, where $N_E$ is the particle number operator of the environment, $\mu$ is the fixed chemical potential of the environment, and $\Xi_E(\beta,\mu)$ is the grand partition function. Accordingly, all preceding derivations remain valid upon replacing $H_E$ with $H_E - \mu N_E$. In particular, the definition of the Gibbs projection inverse temperature in Eq.~\eqref{eq:def_eff_beta_quantum} is obtained by the same substitution.

This decomposition not only provides an estimate for the accessible range of each axis but also enables the derivation of upper and lower bounds on entropy production when combined with the quantum speed limit.
From Eqs.~\eqref{eq:dif_entropy}, ~\eqref{eq:def_mutual_info} and
$S(\rho_{SE}(t))=S(\rho_{SE}(0))$, we obtain
\begin{align}
    &\Delta D_{Y}+ \Delta D_{Z}\nonumber\\
    &=[S(\rho_S(\tau))-S(\rho_S(0))]+[S(\gamma_{E}(\beta^{\ast}(\tau)))-S(\gamma_{E}(\beta^{\ast}(0)))].
    \label{eq:YZ_quantum}
\end{align}
We define $\Phi_Q(t):=D_{X}(t)+ S(\gamma_{E}(\beta^{\ast}(t)))\geq 0$, which can be evaluated given only the reference temperature and mean energy, where the second term is bounded within $[0, \;\ln d_E]$. In fact, the difference $\Delta \Phi_Q$ is given by 
\begin{align}
    \Delta \Phi_Q=\beta(\mathrm{tr}_{E}[H_E \rho_E(\tau)]-\mathrm{tr}_{E}[H_E \rho_E(0)]).
    \label{eq:phi_relation}
\end{align}
The proof is shown in Appendix~\ref{sec:phi_relation}.
Combining Eqs.~\eqref{eq:EP_decomposition_diff} and~\eqref{eq:YZ_quantum} with Eq.~\eqref{eq:FA_bound} yields
\begin{align}
    &|\Sigma(\tau)-\Delta \Phi_Q|=|S(\rho_S(\tau))-S(\rho_S(0))|\nonumber\\
    &\le \mathcal{T}(\rho_S(\tau),\rho_S(0)) \ln (d_S-1) +h_{b}(\mathcal{T}(\rho_S(\tau),\rho_S(0))) .
    \label{eq:ub_trace_dist}
\end{align}

Suppose that an upper bound $\mathcal{T}(\rho_S(\tau),\rho_S(0))\le \Lambda_{Q}$ is given in terms of physical quantities. Let $\dblbrace{A(t)}:=\sqrt{\mathrm{tr}[A(t)^2\rho_{SE}(t)]-\mathrm[A(t)\rho_{SE}(t)]^2}$ be the standard deviation of the Hermitian operator $A$. Following the quantum speed limit framework introduced in Ref.~\cite{PhysRevResearch.3.023074}, take $\Lambda_Q$ to be
\begin{align}
    \Lambda_Q:=\sin\left(\int_0^{\tau}\dblbrace{H_S(t)+H_{SE}(t)} \ dt\right),
    \label{eq:def_lambda_Q}
\end{align}
for $0\le \int_0^{\tau}\dblbrace{H_S(t)+H_{SE}(t)} \ dt \le \pi / 2$.
The detailed derivation of Eq.~\eqref{eq:def_lambda_Q} is provided in Appendix~\ref{sec:ub_quantum}.
When combining this definition with Eq.~\eqref{eq:ub_trace_dist}, we obtain both upper and lower bounds on entropy production:
\begin{align}
    &|\Sigma(\tau)-\Delta \Phi_Q| \nonumber\\
    &\le 
    \begin{cases}
        \Lambda_Q \ln (d_S-1) +h_{b}(\Lambda_Q) & (\Lambda_Q\le \frac{1}{2})\\
        \Lambda_Q \ln (d_S-1) + \ln 2 & (\frac{1}{2} < \Lambda_Q\le 1).
    \end{cases}
    \label{eq:EP_bound_SL}
\end{align}
This inequality is guaranteed since the binary entropy function $h_{b}(x)$ is monotonically increasing for $x \le 1/2$ and attains its maximum value of $\ln 2$ at $x = 1/2$.
This constitutes the second main result of this paper.
Moreover, the bound vanishes as $\mathcal{T}(\rho_S(\tau),\rho_S(0))\rightarrow 0$, thereby capturing the correct small-distance behavior.

\subsection{Classical Markov jump process}

Consider a continuous-time Markov jump process with $n$ discrete states, $\{B_1, B_2, \dots, B_n\}$. Let $W_{ij}$ denote the time-independent transition rate from state $B_j$ to state $B_i$, and let $p_i(t)$ be the probability of the system being in state $B_i$ at time $t$. The time evolution of the probability distribution is governed by the master equation:
\begin{align}
    \dot{p}_i(t)&=\sum_{j} W_{ij}p_j(t),
    \label{eq:master_eq_Markov}
\end{align}
where the diagonal elements satisfy $W_{ii} = -\sum_{j, (j \neq i)} W_{ji}$ to ensure probability conservation. 
Consider a system in contact with a single thermal bath at inverse temperature $\beta$, where the following local detailed balance condition is satisfied:
\begin{align}
    \frac{W_{ij}}{W_{ji}}=e^{-\beta(\epsilon_{i}-\epsilon_{j})},
    \label{eq:def_ldb}
\end{align}
where $\epsilon_i$ denotes the energy level associated with state $B_i$.
Under this assumption, the total entropy production up to time $\tau$ is given by
\begin{align}
    \Sigma(\tau)&=\int_0^\tau \sum_{i\neq j} W_{ij}p_j(t)\ln \frac{W_{ij}p_j(t)}{W_{ji}p_i(t)} \ dt .
    \label{eq:def_EP_Markov}
\end{align}
In the classical Markovian limit, where the dynamics are described solely by the system's state, the system--environment correlation contribution vanishes. Consequently, the component associated with total correlations, corresponding to the $Z$-axis, is absent, leaving only the $X$- and $Y$-components, as illustrated in Fig.~\ref{fig:geometry_EP_C}.
In analogy with Eqs.~\eqref{eq:def_X_dist} and~\eqref{eq:def_Y_dist}, we define
\begin{align}
    D^{\mathrm{KL}}_{X}(t) &:= D_{\mathrm{KL}}(G(\beta^{\ast}(t))\Vert G(\beta)), \label{eq:def_X_dist_C}
    \\
    D^{\mathrm{KL}}_{Y}(t) &:= D_{\mathrm{KL}}(P(t)\Vert G(\beta^{\ast}(t))), \label{eq:def_Y_dist_C}
\end{align}
where $\beta^\ast(t)$ is implicitly determined by Eq.~\eqref{eq:def_eff_beta_classic}.
We now discuss the properties of the $X$- and $Y$-axes in turn. The $X$-axis also retains Property 1 in the classical case. It follows that Eq.~\eqref{eq:def_X_dist_C} takes the same form as in the open quantum case.
Let $d^{\mathrm{TV}}_{Y}(t):=d_{\mathrm{TV}}(P(t), G(\beta^{\ast}(t)))$.
As in Eq.~\eqref{eq:lu_bound_Y}, the range of the $Y$-axis is bounded by the total variation distance as follows:
 \begin{align}
    2d^{\mathrm{TV}}_{Y}(t)^2 &\le D^{\mathrm{KL}}_{Y}(t) \le d^{\mathrm{TV}}_{Y}(t)\ln(n-1)+h_{b}\left(d^{\mathrm{TV}}_{Y}(t)\right).
    \label{eq:lu_bound_Y_C}
\end{align}
This inequality can be proven in a manner similar to the open quantum case.

We next derive the upper and lower bounds on the entropy production using the speed limit, in exact analogy with the quantum case.
Combining Eqs.~\eqref{eq:def_ldb} and~\eqref{eq:def_EP_Markov}, and applying Eq.~\eqref{eq:Pythagorean_eff}, the entropy production can be written as
\begin{align}
    &\Sigma(\tau)=H_{\mathrm{SH}}(P(\tau))-H_{\mathrm{SH}}(P(0))-\beta(\braket{\epsilon(\tau)}-\braket{\epsilon(0)})\nonumber\\
    &=D_{\mathrm{KL}}(P(0)\Vert G(\beta))-D_{\mathrm{KL}}(P(\tau)\Vert G(\beta))\nonumber\\
    &=-\Delta D^{\mathrm{KL}}_{X}-\Delta D^{\mathrm{KL}}_{Y},
\end{align}
where $\braket{\epsilon(t)}:=\sum_i \epsilon_i p_i(t)$.
Equation~\eqref{eq:dif_entropy} reduces to
\begin{align}
    &\Delta D^{\mathrm{KL}}_Y(\tau)=-[H_{\mathrm{SH}}(P(\tau))-H_{\mathrm{SH}}(P(0))]\nonumber\\
    &+[H_{\mathrm{SH}}(G(\beta^{\ast}(\tau))-H_{\mathrm{SH}}(G(\beta^{\ast}
    (0))].
\end{align}
Since this is a similar expression to its quantum counterpart [Eq.~\eqref{eq:YZ_quantum}], combining $\Sigma(\tau)=-\Delta D^{\mathrm{KL}}_{X}-\Delta D^{\mathrm{KL}}_{Y}$ with Eq.~\eqref{eq:FA_bound_classic} yields
\begin{align}
    &|\Sigma(\tau)+\Delta \Phi_C|=|H_{\mathrm{SH}}(P(\tau))-H_{\mathrm{SH}}(P(0))|\nonumber\\
    &\le d_{\mathrm{TV}}(P(\tau),P(0)) \ln (n-1) +h_{b}(d_{\mathrm{TV}}(P(\tau),P(0))) .
    \label{eq:ub_TV_dist}
\end{align}
where $\Phi_C(t):=D^{\mathrm{KL}}_{X}(t)+ H_{\mathrm{SH}}(G(\beta^{\ast}(t)))\geq 0$. As in Eq.~\eqref{eq:phi_relation}, the relation $\Delta \Phi_C=\beta(\braket{\epsilon(\tau)}-\braket{\epsilon(0)})$ holds.
Letting $d_{\mathrm{TV}}(P(\tau),P(0))\le \Lambda_{C}$, from the result in Ref.~\cite{Hasegawa:2023:BulkBoundaryBoundNC}, we take $\Lambda_{C}$ to be
\begin{align}
    \Lambda_C:=\sin\left(\frac{1}{2}\int_0^\tau \frac{\sqrt{\mathcal{A}(t)}}{t}\ dt\right),
    \label{eq:def_lambda_C}
\end{align}
for $0 \le \int_0^\tau \sqrt{\mathcal{A}(t)/t}\ dt \le \pi $.
Here, $\mathcal{A}(t):=\int_0^\tau \sum_{i\neq j} W_{ij}p_j(t)\,dt$ is the dynamical activity, which quantifies the activity of systems by the average number of jump events during $[0,\tau]$.
The detailed derivation of Eq.~\eqref{eq:def_lambda_C} is provided in Appendix~\ref{sec:ub_classic}.
Since Eqs.~\eqref{eq:ub_trace_dist} and~\eqref{eq:ub_TV_dist} share the same functional form, we can follow a procedure identical to that used for Eq.~\eqref{eq:EP_bound_SL} to obtain the following bounds:
\begin{align}
    &|\Sigma(\tau)+\Delta \Phi_C| \nonumber\\
    &\le 
    \begin{cases}
        \Lambda_C \ln (n-1)+h_{b}(\Lambda_C) & (\Lambda_C\le \frac{1}{2})\\
        \Lambda_C \ln (n-1) + \ln 2 & (\frac{1}{2} < \Lambda_C\le 1).
    \end{cases}
    \label{eq:EP_bound_SL_classic}
\end{align}
This is the third main result of this paper.
While a previous study in Ref.~\cite{Nishiyama:2023:EPUpperBound} derived an upper bound using $\max_{i,j} |\epsilon_i-\epsilon_j|$, Eq.~\eqref{eq:EP_bound_SL_classic} provides both upper and lower bounds solely in terms of the average energy and the dynamical activity.

\section{Example}

To demonstrate the decomposition shown in Eq.~\eqref{eq:EP_decomposition_diff},
we consider the noninteracting resonant model, which comprises a fermionic system coupled to a fermionic bath.
This model was employed in Ref.~\cite{Ptaszynski:2022:PostThermalization} to study the behavior of quantum entropy production.
The total Hamiltonian $H$ comprises $H_S$ (system Hamiltonian), $H_E$ (bath Hamiltonian), and $H_I$ (interaction Hamiltonian):
\begin{align}
    H_{S}&=\epsilon_{d}c^{\dagger}_{d}c_{d},\label{eq:HS_def}\\H_{E}&=\sum_{k}\epsilon_{k}c^{\dagger}_{k}c_{k},\label{eq:HE_def}\\H_{I}&=\sum_{k}\left(\Omega c^{\dagger}_{d}c_{k}+\mathrm{h.c.}\right),\label{eq:HI_def}
\end{align}
Here, $c_d$ is the annihilation operator.
$\epsilon_k$ is the energy level of the fermionic bath.
$\Omega$ is the tunnel coupling strength.

\begin{figure}
\centering
\includegraphics[width=1\linewidth]{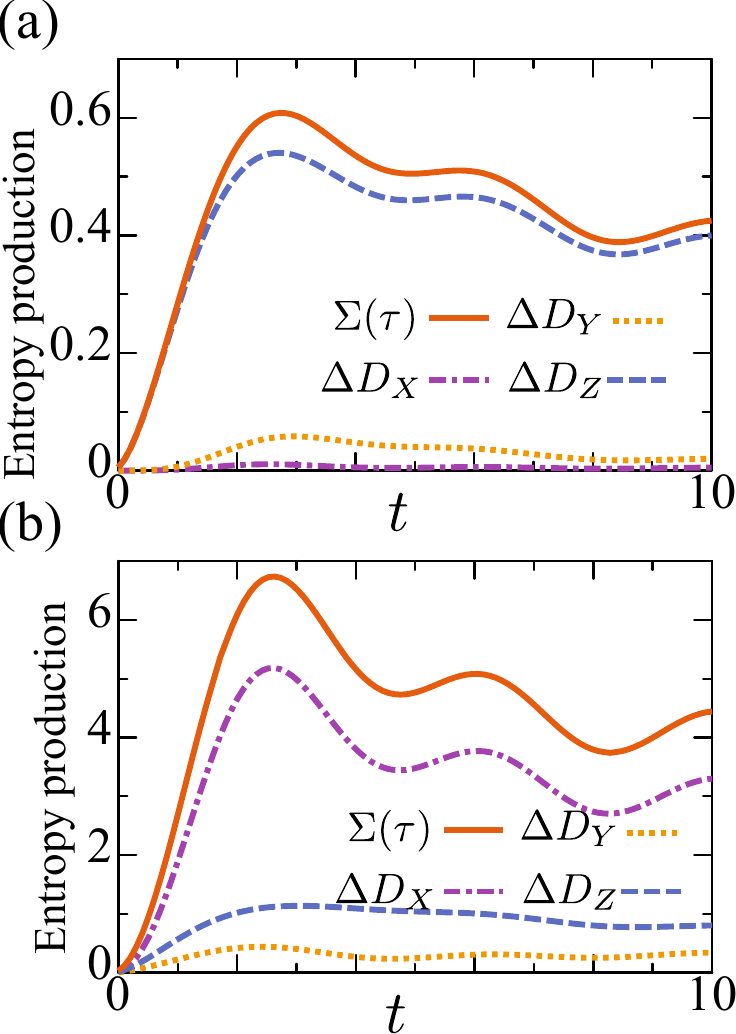}
\caption{Decomposition of the entropy production given in Eq.~\eqref{eq:EP_decomposition_diff} for (a) $\beta=1$ and (b) $\beta=30$. 
The other settings are $\mu = 1$,
$K = 5$,
$W = 1$, and 
$\epsilon_d = 1$. 
$\Sigma(\tau)$, $\Delta D_X$, $\Delta D_Y$, and $\Delta D_Z$ are represented by the solid, dot-dashed, dotted, and dashed lines, respectively, as functions of time $t$. 
}
\label{fig:entropy_production}
\end{figure}

We perform numerical simulations for the model described by Eqs.~\eqref{eq:HS_def}-\eqref{eq:HI_def}. 
Figure~\ref{fig:entropy_production} shows the time evolution of entropy production and its components $\Delta D_X$, $\Delta D_Y$, and $\Delta D_Z$ for (a) $\beta=1$ and (b) $\beta=30$.
The other parameter settings are described in the caption of Fig.~\ref{fig:entropy_production}. 
Here, the solid line indicates the entropy production, the dash-dotted line the $X$-component, the dotted line the $Y$-component, and the dashed line the $Z$-component.
For the higher-temperature case (Fig.~\ref{fig:entropy_production}(a)),
evidently, the $Z$-component, representing the correlation between the primary system and the environment, is the dominant contribution.
On the other hand, for the lower-temperature case (Fig.~\ref{fig:entropy_production}(b)), the $X$-component exhibits the largest contribution to the entropy production.

\section{Conclusion}

In conclusion, we have established a geometric framework for entropy production based on a hierarchical projection structure of quantum relative entropy. While decompositions of entropy production into physically distinct contributions have been considered in previous studies, our approach reveals the geometric origin of these contributions and, more importantly, the intrinsic anisotropy underlying their different mathematical inequalities and information-theoretic constraints.

By exploiting the Pythagorean identity of quantum relative entropy, we derived an exact orthogonal decomposition into three contributions associated with the deviation of the Gibbs projection temperature from the reference temperature, the deviation of the environment from its Gibbs projection state, and system--environment correlations. The significance of this decomposition is not merely the separation of entropy production into individual terms, but the fact that each contribution is subject to a distinct geometric constraint. This anisotropic structure accounts for the fundamentally different constraints obeyed by the individual contributions: the environmental deviation and correlation contributions admit trace-distance-based bounds, whereas the Gibbs projection temperature contribution is governed by a classical Kullback--Leibler divergence and admits no universal upper bound in terms of the trace distance and system and environment dimensions alone.

Combining the geometric decomposition with quantum speed limits, we obtained physically accessible upper and lower bounds on entropy production in terms of mean energies and Hamiltonian variances. The same framework also extends to classical Markovian dynamics described solely by the system's state, where the correlation contribution is absent from the decomposition and the geometric structure reduces to a classical two-component decomposition characterized by the total variation distance.

These results demonstrate that the hierarchical projection structure provides a unifying geometric origin for the diverse features of entropy production, connecting information geometry, nonequilibrium thermodynamics, and dynamical speed limits. More broadly, our findings suggest that anisotropic geometric structures may provide a useful perspective for understanding the distinct contributions to irreversibility in quantum and classical systems.

\begin{acknowledgements}

This work was supported by JSPS KAKENHI Grant Numbers JP26K02998, JP24K03008, and JP26K06158.

ChatGPT (free plan) and Gemini Flash were used to assist with code development and text preparation of this manuscript. 
All scientific ideas, analyses, calculations, and conclusions are those of the authors.
\end{acknowledgements}

\appendix
\begin{widetext}

\section{Proof of the Pythagorean identity [Eq.~\eqref{eq:Pythagorean_ID}] \label{sec:pythagorean}}
\begin{align}
    &D(\rho\Vert\omega)=\mathrm{tr}[(\rho(\ln\rho-\ln\omega)]=\mathrm{tr}[\rho(\ln\rho-\ln\sigma)]+\mathrm{tr}[\rho(\ln\sigma-\ln\omega)]\nonumber\\
    &=D(\rho\Vert \sigma)+\mathrm{tr}[\sigma(\ln\sigma-\ln\omega)]
    =D(\rho\Vert \sigma)+D(\sigma\Vert \omega),
\end{align}
where we use the orthogonal condition [Eq.~\eqref{eq:orthogonal_cond}] in the third equality.

\section{Uniqueness of solution of Eq.~\eqref{eq:def_eff_beta} \label{sec:uniqueness}}
Differentiating $\mathbb{E}_\gamma[H](\beta) := \mathrm{tr}[H\gamma(\beta) ]$ with respect to $\beta$ yields
\begin{align}
    \frac{d}{d\beta}\mathbb{E}_\gamma[H](\beta) = -\mathrm{Var}_\gamma[H](\beta) < 0,
    \label{eq:monotonicity}
\end{align}
where $\mathrm{Var}_\gamma[H](\beta) := \mathbb{E}_\gamma[H^2](\beta) - \left( \mathbb{E}_\gamma[H](\beta) \right)^2$ denotes the variance of $H$. Let $\epsilon_{\max}$ and $\epsilon_{\min}$ be the maximum and minimum eigenvalues of $H$, respectively. Since $\mathbb{E}_\gamma[H](\beta)$ is strictly decreasing with $\mathbb{E}_\gamma[H](-\infty) = \epsilon_{\max}$ and $\mathbb{E}_\gamma[H](\infty) = \epsilon_{\min}$, Eq.~\eqref{eq:def_eff_beta} always possesses a unique solution.

\section{Proof of properties \label{sec:proof_properties}}

\subsection{Proof of property 1}

Since $\gamma_E(\beta^{\ast}(t))$ commutes at any time $t$, we obtain $D_{X}(t) = D_{\mathrm{KL}}(G_E(\beta^{\ast}(t)) \Vert G_E(\beta))$.

Assume the existence of a function $f$ such that $D_{X}(t)=D(\gamma_E(\beta^{\ast}(t))\Vert\gamma_E(\beta))\le f(\mathcal{T}_{X}(t); \;d_S, d_E)$. Let us consider an environment modeled as a two-level system with eigenvalues $\epsilon_{\min}=0$ and $\epsilon_{\max}=\epsilon>0$ (with eigenstates $\ket{g}$ and $\ket{e}$). The initial state for a given inverse temperature $\beta$ is given by
\begin{align}
    \rho_E(0) = \gamma_E(\beta)= \frac{1}{1 + e^{-\beta \epsilon}} \lvert g \rangle \langle g \rvert + \frac{e^{-\beta \epsilon}}{1 + e^{-\beta \epsilon}} \lvert e \rangle \langle e \rvert.
\end{align}
By combining $D_{X}(t)=-S(\gamma_{E}(\beta^{\ast}(t)))+\beta\mathrm{tr}_{E}[H_{E} \gamma_{E}(\beta^{\ast}(t))]+\ln \mathcal{Z}(\beta)$ with $\ln \mathcal{Z}(\beta(t))\geq 0$ and $S(\gamma_E(\beta^{\ast}(t)))\le \ln d_E$, it follows that
\begin{align}
    \beta\mathrm{tr}[H_E \gamma_E(\beta^{\ast}(t))]-\ln d_E \le D_{X}(t) \le f(\mathcal{T}_{X}(t); \;d_S, d_E).
    \label{eq:upperbound_X}
\end{align}
Let $\sigma_{E}(0)=\ket{g}\bra{g}$.
For a fixed short evolution time $t=\Delta t$, the environment's density matrix $\sigma_{E}(\Delta t)$ can be parameterized as
\begin{align}
    \sigma_E(\Delta t) = (1-\delta)\lvert g \rangle\langle g \rvert + \delta\lvert e \rangle\langle e \rvert + \left( c\ket{g} \bra{e} + c^* \ket{e}\bra{g}\right),
\end{align}
where $\delta:=\braket{e|\sigma_E(\Delta t)|e}$ is the excited-state population at 
 $t=\Delta t$, and $c \in \mathbb{C}$ accounts for the induced coherence satisfying $\vert{}c\vert{}^2 \le \delta(1-\delta)$.
We can assume that the excited-state population at $t=\Delta t$ remains small, $\delta \ll 1$.
In the low-temperature limit $\beta \epsilon \gg 1$, the initial thermal population of the excited state is exponentially suppressed. Hence, we obtain
\begin{align}
    \rho_E(\Delta t) = \sigma_{E}(\Delta t) +\mathcal{O}(e^{-\beta\epsilon}),
\end{align}
Under this evolution, the corresponding Gibbs state at $\beta^{\ast}(\Delta t)$ reduces to
\begin{align} 
    \gamma_E(\beta^{\ast}(\Delta t)) = (1-\delta)\lvert g \rangle\langle g \rvert + \delta\lvert e \rangle\langle e \rvert+\mathcal{O}(e^{-\beta\epsilon}).
\end{align}
Since the trace distance scales as $\mathcal{T}_{X}(\Delta t) = \delta + \mathcal{O}(e^{-\beta\epsilon}) \ll 1$, Eq.~\eqref{eq:upperbound_X} can be recast into the form
\begin{align}
    \beta\epsilon\delta - \ln d_{E} \le f\big(\delta; d_S, d_E\big) + \mathcal{O}(e^{-\beta\epsilon}).
\end{align}
Note that the product $\epsilon \delta$ remains fixed, whereas the inverse temperature $\beta$ can be taken to be arbitrarily large. Taking the limit $\beta \to \infty$ on the left-hand side leads to a clear contradiction, as the upper bound on the right-hand side remains bounded. Therefore, such a universal function $f$ cannot exist, which rigorously demonstrates that $D_{X}(t)$ can become arbitrarily large even in the regime where $\mathcal{T}_{X}(t) \ll 1$.

\subsection{Proof of property 2}

We establish the upper bound in Eq.~\eqref{eq:lu_bound_Y}. The upper bound follows from Eqs.~\eqref{eq:dif_entropy} and~\eqref{eq:FA_bound}. Combined with the entropic bound $0 \le S(\sigma_E) \le \ln d_{E}$ for any density operator $\sigma_E$, Eq.~\eqref{eq:dif_entropy} readily yields $D_{Y}(t) \le \ln d_E$. 

\subsection{Proof of property 3}

We show the proof of the upper bound in Eq.~\eqref{eq:lu_bound_Z}.
Let $S(A|B)_{\rho}:=S(\rho_{AB})-S(\rho_{B})$ be the conditional entropy. The Alicki–Fannes–Winter inequality~\cite{winter2016tight} states that
\begin{align}
    \left|S(A|B)_{\rho}-S(A|B)_{\sigma}\right|\le 2\mathcal{T}(\rho_{AB}, \sigma_{AB})\ln d_{A}+g\left(\mathcal{T}(\rho_{AB}, \sigma_{AB})\right).
    \label{eq:AFW_ineq}
\end{align}
Let $A=S$ and $B=E$.
Since the mutual information can be written as
\begin{align}
    D_Z(t) = I_{S:E}(t) = S(S|E)_{\rho_{S}(t)\rho_{E}(t)}-S(S|E)_{\rho_{SE}(t)},
\end{align}
substituting $\rho_{AB} = \rho_{S}(t)\rho_{E}(t)$ and $\sigma_{AB} = \rho_{SE}(t)$ into Eq.~\eqref{eq:AFW_ineq} yields the following upper bound:
\begin{align}
    D_Z(t)=|S(S|E)_{\rho_{S}(t)\rho_{E}(t)}-S(S|E)_{\rho_{SE}(t)}|\le 2\mathcal{T}_{Z}(t)\ln d_{S}+g\left(\mathcal{T}_{Z}(t)\right).
    \label{eq:Dz_ub}
\end{align}
Since the same inequality holds upon exchanging $S$ and $E$, we obtain the upper bound in Eq.~\eqref{eq:lu_bound_Z}.
From the result in Ref.~\cite{araki1970entropy}, $D_{Z}(t)=I_{S:E}(t)\le 2\min(S(\rho_{S}(t)), S(\rho_{E}(t)))\le 2\min(\ln d_{S}, \ln d_{E})$ follows.

\section{Proof of Eq.~\eqref{eq:phi_relation} \label{sec:phi_relation}}

From the definition of $\Phi_Q(t)$ and Eq.~\eqref{eq:def_X_dist}, it follows that
\begin{align}
    \Phi_Q(t)=D_{X}(t)+ S(\gamma_{E}(\beta^{\ast}(t)))=\beta\mathrm{tr}_{E}[H_{E} \gamma_{E}(\beta^{\ast}(t))]+\ln\mathcal{Z}(\beta)=\beta\mathrm{tr}_{E}[H_{E} \rho_{E}(t)]+\ln\mathcal{Z}(\beta),
\end{align}
where we use Eq.~\eqref{eq:def_eff_beta_quantum}.
Therefore, we obtain Eq.~\eqref{eq:phi_relation}.

\section{Proof of Eq.~\eqref{eq:def_lambda_Q} \label{sec:ub_quantum}}

Let $\mathcal{L}_D(\rho, \sigma)$ be the Bures angle~\cite{Nielsen:2011:QuantumInfoBook} defined as follows:
\begin{align}
    \mathcal{L}_D(\rho, \sigma):= \arccos\left[\sqrt{\mathrm{Fid}(\rho, \sigma)}\right],
    \label{eq:L_D_def}
\end{align}
where $\mathrm{Fid}(\rho, \sigma)$ is the quantum fidelity: 
\begin{align}
    \mathrm{Fid}(\rho,\sigma):=\left(\mathrm{Tr}\left[\sqrt{\sqrt{\rho}\sigma\sqrt{\rho}}\right]\right)^{2}.
    \label{eq:fidelity_def}
\end{align}
From the result in Ref.~\cite{PhysRevResearch.3.023074}, the quantum speed limit is given by
\begin{align}
    \mathcal{L}_D(\rho_{S}(\tau), \rho_{S}(0))\le \int_0^{\tau}\dblbrace{H_S(t)+H_{SE}(t)} \ dt.
\end{align}
Combining this inequality with $\mathcal{T}(\rho,\sigma)\le \sqrt{1-\mathrm{Fid}(\rho,\sigma)}$~\cite{fuchs1999cryptographic} yields Eq.~\eqref{eq:def_lambda_Q}.

\section{Proof of Eq.~\eqref{eq:def_lambda_C} \label{sec:ub_classic}}

Let $\mathcal{L}_P(P,Q)$ be the Bhattacharyya angle: 
\begin{align}
    \mathcal{L}_P(P,Q):=\arccos\left(BC(P,Q)\right),
    \label{eq:Bhattacharyya_arccos_def}
\end{align}
where $BC(P,Q):=\sum_i \sqrt{p_i q_i}$ is the Bhattacharyya coefficient.
From the result in Ref.~\cite{Hasegawa:2023:BulkBoundaryBoundNC}, it follows that
\begin{align}
    \mathcal{L}_P(P(\tau),P(0))\le \frac{1}{2}\int_0^\tau \frac{\sqrt{\mathcal{A}(t)}}{t} \ dt.
\end{align}    
Combining these relations with $d_{\mathrm{TV}}(P,Q)\le \sqrt{1-BC(P,Q)^2}$ yields Eq.~\eqref{eq:def_lambda_C}.

\section{Multiple heat baths \label{sec:multi_bath}}

Let us consider an open quantum system coupled to multiple heat baths labeled by $\alpha \in {1,2,\ldots ,N}$.
The environmental Hamiltonian is assumed to be
\begin{align}
    H_E=\sum_{\alpha}H_{\alpha}.
\end{align}
The reduced state of the $\alpha$-th bath at time $t$ is
\begin{align}
    \rho_{\alpha}(t):=\mathrm{tr}_{E\setminus{\alpha}}[\rho_E(t)],
\end{align}
where $\mathrm{tr}_{E\setminus{\alpha}}$ denotes the partial trace over all environmental degrees of freedom except the $\alpha$-th bath. 
In this section, $\beta_{\alpha}$ denotes the inverse temperature parameter specifying the reference Gibbs state $\gamma_{\alpha}(\beta_{\alpha})$ of the $\alpha$-th bath. The state $\gamma_{\alpha}(\beta_{\alpha})$ is defined in Eq.~\eqref{eq:def_Gibbs_state} associated with the Hamiltonian $H_{\alpha}$.
Let $G_\alpha(\beta_\alpha) := \{\mathfrak{g}_{i,\alpha}(\beta_\alpha)\}$ be the Gibbs distribution for the $\alpha$-th bath defined in Eq.~\eqref{eq:def_Gibbs_dist}, where the index $i$ labels the energy levels of the $\alpha$-th bath.
We then define the Gibbs projection inverse temperature of the $\alpha$-th bath through
\begin{align}
    \mathrm{tr}_{\alpha}[H_\alpha \gamma_{\alpha}(\beta^{\ast}(t))]=\mathrm{tr}_{\alpha}[H_{\alpha}\rho_{\alpha}(t)],
    \label{eq:def_eff_beta_quantum_multi}
\end{align}
where $\mathrm{tr}_{\alpha}$ denotes the partial trace over the $\alpha$-th bath.
The entropy production is then expressed as
\begin{align}
    &\Sigma(\tau) = D\left(\rho_{SE}(\tau)\Vert\rho_{S}(\tau)\prod_{\alpha}\gamma_{\alpha}(\beta_{\alpha})\right)- D\left(\rho_{SE}(0)\Vert\rho_{S}(0)\prod_{\alpha}\gamma_{\alpha}(\beta_{\alpha})\right).
    \label{eq:EP_multi_bath}
\end{align}
To align with the axis definitions given in Eqs.~\eqref{eq:def_X_dist} and \eqref{eq:def_Y_dist}, we replace $D_{X}(t)$ and $D_{Y}(t)$ as follows:
\begin{align}  
    D_{X}(t)&=D\left(\prod_{\alpha}\gamma_{\alpha}(\beta^{\ast}_{\alpha}(t))\Vert \prod_{\alpha}\gamma_{\alpha}(\beta_{\alpha})\right) =\sum_{\alpha} D_{\mathrm{KL}}(G_{\alpha}(\beta^{\ast}_{\alpha}(t))\Vert G_{\alpha}(\beta_{\alpha})), \label{eq:def_X_dist_multi}\\
    D_{Y}(t)&=D\left(\rho_{E}(t)\Vert \prod_{\alpha}\gamma_{\alpha}(\beta^{\ast}_{\alpha}(t))\right)=D_{cor}(t)+\sum_{\alpha}D_{\alpha}(t). \label{eq:def_Y_dist_multi}
\end{align}
Here, we define
\begin{align}
    D_{cor}(t):=D\left(\rho_{E}(t)\Vert \prod_{\alpha}\rho_{\alpha}(t)\right),
    \label{eq:def_cor_dist}
\end{align}
and 
\begin{align}
    D_{\alpha}(t):=D\left(\rho_{\alpha}(t)\Vert \gamma_{\alpha}(\beta^{\ast}_{\alpha}(t))\right).
    \label{eq:def_alpha_dist}
\end{align}
The sub-component $D_{\text{cor}}(t)$ accounts for the correlations among the multiple baths.
In Eq.~\eqref{eq:def_Y_dist_multi}, we apply Eq.~\eqref{eq:orthogonal_cond} with $\rho=\rho_{E}(t)$, $\sigma=\prod_{\alpha}\rho_{\alpha}(t)$, and $\omega=\prod_{\alpha}\gamma_{\alpha}(\beta^{\ast}_{\alpha}(t))$.
For the $Z$-axis, the same definition as in Eq.~\eqref{eq:def_Z_dist} is adopted.

We show below that these quantities satisfy the decomposition given by Eq.~\eqref{eq:EP_decomposition_diff}.
As in Eq.~\eqref{eq:def_Y_dist_multi}, it follows that 
\begin{align}
    &D\left(\rho_{E}(t)\Vert \prod_{\alpha}\gamma_{\alpha}(\beta_{\alpha})\right)
    =D_{cor}(t)+\sum_{\alpha}D\left(\rho_{\alpha}(t)\Vert\gamma_{\alpha}(\beta_{\alpha})\right)
    \nonumber\\
    &=D_{cor}(t)+\sum_{\alpha}D \left(\rho_{\alpha}(t)\Vert\gamma_{\alpha}(\beta^{\ast}_{\alpha}(t))\right)+\sum_{\alpha}D_{\mathrm{KL}}\left(G_{\alpha}(\beta^{\ast}_{\alpha}(t))\Vert G_{\alpha}(\beta_{\alpha})\right)
    =D_{X}(t)+D_{Y}(t).
\end{align}
where we use Eq.~\eqref{eq:Pythagorean_eff} in the second equality.
Using this relation, Eqs.~\eqref{eq:EP_decomp_old} and~\eqref{eq:EP_multi_bath}, and the definition of $D_Z(t)$ [Eq.~\eqref{eq:def_Z_dist}], we obtain the following decomposition:
\begin{align}
    \Sigma(\tau)=\Delta D_{X}+\Delta D_{Y}+\Delta D_{Z}.
\end{align}
This recovers the same expression as Eq.~\eqref{eq:EP_decomposition_diff}.

Along the $X$-axis, $D_{X}(t)$ represents the classical contribution defined in Eq.~\eqref{eq:def_X_dist_multi}. As implied by Property 1, it admits no upper bound in terms of the trace distance. We now show that, as in the single-bath case, the contribution along the $Y$-axis is bounded by the trace distance and the dimension of the environment, demonstrating that the intrinsic anisotropy persists in the multi-bath setting.
Let $\mathcal{T}_{\alpha}(t):= \mathcal{T}\left(\rho_{\alpha}(t), \gamma_{\alpha}(\beta^{\ast}_{\alpha}(t))\right)$, and let $d_{\alpha}$ denote the dimension of the $\alpha$-th bath. Then, as in Eq.~\eqref{eq:lu_bound_Y}, $D_{\alpha}(t)$ is bounded as
\begin{align}
    &2 \mathcal{T}_{\alpha}(t)^2 \le D_{\alpha}(t) \le \mathcal{T}_{\alpha}(t)\ln(d_{\alpha}-1)+h_{b}\left(\mathcal{T}_{\alpha}(t)\right).
    \label{eq:lu_bound_Y_alpha}
\end{align}
We next derive the upper and lower bounds for $D_{cor}(t)$. Let $\mathcal{T}_{cor}(t):= \mathcal{T}\left(\rho_{E}(t), \; \prod_{\alpha}\rho_{\alpha}(t)\right)$. 
While the lower bound $2\mathcal{T}_{cor}(t)^2\le D_{cor}(t)$ directly follows from the quantum Pinsker inequality, here we focus on establishing the upper bound. From Eq.~\eqref{eq:def_cor_dist}, we obtain
\begin{align}
    D_{cor}(t)=\sum_{\alpha} S(\rho_{\alpha}(t)) -S(\rho_{E}(t))=S\left(\prod_{\alpha}\rho_{\alpha}(t)\right) -  S(\rho_{E}(t)).
\end{align}
Applying the the Fannes–Audenaert inequality [Eq.~\eqref{eq:FA_bound}] yields
\begin{align}
    &2\mathcal{T}_{cor}(t)^2\le D_{cor}(t)\le \mathcal{T}_{cor}(t)\ln (d_{E}-1) + h_b(\mathcal{T}_{cor}(t)).
    \label{eq:lu_bound_cor_multi}
\end{align}

\end{widetext}

\end{document}